\documentclass[a4paper,11pt]{article}
\usepackage{pos}
\usepackage{amsmath}
\usepackage{amscd}
\usepackage{amsfonts}
\usepackage{mathrsfs}
\usepackage{fancybox} 
\usepackage{enumerate}
\usepackage[boxsize=0.8ex]{ytableau} %young diagram
\usepackage{youngtab}
\usepackage{graphicx}
\usepackage{url} 
\usepackage{color}
\usepackage{ulem}
\usepackage{comment}

\DeclareMathOperator{\tr}{\mathrm{tr}}

\def\={\stackrel{\bullet}{=}}

\DeclareMathOperator{\Tr}{\mathrm{Tr}}
\def\z#1{\mathbb{Z}_{#1}} % Z_{} Symmetry
\def\zn#1{$\mathbb{Z}_{#1}$} % 文中　Z_{} Symmetry

\def\mscr#1{\mathscr{#1}}
\def\mcl#1{\mathcal{#1}}
\def\tx#1{\text{#1}}

\def\({\left(}
\def\){\right)}
\def\[{\left[}
\def\]{\right]}

\renewcommand{\hookAfterAbstract}{%
  \par\bigskip
  \textsc{Preprint Number}: OU-HET-1262%
}

\title{SU(6) model revisited}
\author[*]{Tetsuya Onogi, Hiroki Wada, Tatsuya Yamaoka}

\affiliation[]{Department of Physics, Osaka University, Toyonaka, Osaka 560-0043, Japan}

\emailAdd{onogi@het.phys.sci.osaka-u.ac.jp}
\emailAdd{hwada@het.phys.sci.osaka-u.ac.jp}
\emailAdd{t\_yamaoka@het.phys.sci.osaka-u.ac.jp}

\abstract{
    We discuss the vacuum structure of the SU(6) model, 
    a chiral gauge theory, from the perspective of anomaly matching. 
    To this end, we first identify all possible 't Hooft anomalies in the UV theory 
    using the Stora-Zumino procedure. 
    Subsequently, we construct an effective theory 
    by applying the idea of the Wess-Zumino-Witten action to derive the topological terms 
    that encode the 't Hooft anomalies. 
    As a result, we demonstrate that a low-energy effective theory 
    reproducing one of the anomalies, namely the mixed anomaly, is described by a $\z 3$
 -valued scalar field. On the other hand, the effective theory 
 that accounts for the discrete chiral self-anomaly is significantly more intricate, 
 and elucidating its structure remains an ongoing challenge.
}

\FullConference{The 41st International Symposium on Lattice Field Theory (LATTICE2024)\\
 28 July - 3 August 2024\\
Liverpool, UK\\}

\begin{document}
\maketitle

%%%%%%%%%%%%%%%%%%%%%%%%%%%%%%%%%%%%%%%%%%%%%%%%%%%%
%%%%%%%%%%%%%%%%%%%%%%%%%%%%%%%%%%%%%%%%%%%%%%%%%%%%
\section{Introduction}
Realizing chiral gauge theory on a lattice has been a longstanding problem.
Toward a solution to this problem,
in this report, we focus on the SU(6) model, which is one of the chiral gauge theory.
There are two main reasons why this SU(6) model is interesting.
The first reason is that  it might be realized on the lattice,
because the Weyl fermions in this model follow the self-conjugate representation.
Realizing this model on the lattice could provide hints for realizing all chiral gauge theories on the lattice.
The second is that in terms of anomaly matching, the vacuum structure is highly non-trivial as we will see later.
Indeed, the fermion bilinear condensate cannot be formed when the chiral symmetry is spontaneously broken.
In that sense, we can expect there should exist the non-trivial degrees of freedom in the vacuum.
That is why this model is very interesting.
In order to realize this model on the lattice, it is very important to first understand the vacuum structure.
That is why the main topic in this paper is about the vacuum structure in this model.

Our motivation in this work is to understand the vacuum structure via 't~Hooft anomaly matching conditions including generalized symmetry.
In order to do that, 
we  analyze all the  't~Hooft anomalies which are possible to arise, 
And by using the 't~Hooft anomaly matching conditions about them, we reveal the IR vacuum structures.
And our claim that 
under the assumption 
that the order parameter of spontaneously symmetery breaking (SSB) of chiral symmetry is four-fermi operator,
all the anomalies in this model are captured by only one scalar field in the IR region.  
It implies the vacuum structure consists of one scalar field with three-fold vacua.

%%%%%%%%%%%%%%%%%%%%%%%%%%%%%%%%%%%%%%%%%%%%%%%%%%%%
%%%%%%%%%%%%%%%%%%%%%%%%%%%%%%%%%%%%%%%%%%%%%%%%%%%%
\section{SU(N) gauge theory with self-adjoint chiral fermions} 
We consider the SU(N) gauge theory (N even) with Weyl fermions in the $\frac{N}{2}$ fully antisymmetric representation which is an self-conjugate representation. 
In this section, we shortly review the chiral symmetry breaking of the system due to the 0-form - 1-form mixed 't Hooft anomaly.\cite{Yamaguchi:2018xse,Bolognesi:2019fej}.

\subsection{The symmetries }
\label{the symmetries}
As we will discuss $SU(6)$ models in details later, we limit ourselves to the case where N=6~\footnote{Of course, it is possible to consider extending $N=6$ to the case of $N \neq 6$. 
Indeed, for general N, we can consider center symmetry and discrete chiral symmetry as well as $N=6$.}.
Then the total symmetry which consists of dynamical gauge symmetry $SU(N)$ and the global symmetries of our system is given by
\begin{equation}
\label{gauge group}
    G =  \frac{SU(6)\times \z {l=6}^\chi}{\z 6}\cong \frac{SU(6)/\z {q=3}^c \times \z 6^\chi}{\z 2}\sim \frac{SU(6)}{\z 3^c} \times \z {l=6}^\chi.
\end{equation}
where $\z {q=3}^c$ is the global center symmetry of dynamical color symmetry $SU(6)$, $\z 6^\chi$ is the discrete chiral symmetry, and
$\z 2$ of the denominator in eq.(\ref{gauge group}) is because the $SU(6)/\z 3^c$ is shared by elements in $\z 6^\chi$.
%It is obvious that this $\z 2 ^{c (0)} = SU(6)/ \z {q=3}$ transformation is overlapping with $\z 2 ^\chi \subset \z 6^\chi$. Therefore the total symmetry is given by eq.(\ref{gauge group}).
The third equal "$\sim$" is justified because the 2-form background gauge field of $\z 2$ can be vanished by the 1-form gauge transformation. For more details about it, see \ref{Cancellation of z 2 symmetry}.

Previous works\cite{Yamaguchi:2018xse} and \cite{Bolognesi:2019fej}, considered a portion of total symmetry; they used 't Hooft anomaly matching condition\cite{tHooft:1979rat} of the 1-form center symmetry $\z q ^{(1)}$ and the 0-form chiral symmetry $\z l ^\chi$, and
find constraints on the spontaneously chiral symmetry breaking in confining phase by gauging only part of total group (\ref{gauge group}), i.e., $\z q ^{(1)}$ or $\z q ^{(1)}$ and $\z l ^{\chi(0)}$ 
(Here, the superscript denotes the form order.)\footnote{ 
    Actually, gauging this $\z 2$ is trivial; the back ground gauge field can be trivial by the gauge fixing. We consider its detail in Sec.\ref{Gauging the chiral symmetry} .}.
So it is important to figure out the meaning of each symmetry in our model, so that let us see each symmetry in detail. 
%Therefore, here, we discuss about the chiral symmetry and the center symmetry$\z q ^{(1)}$ .
\begin{table}[htbp]
  \centering
  \begin{tabular}{|c|c|c|c|c|}
    \hline
    rep. & Dynkin index: $l$ & N-ality: $c$ &q=gcd$\{N,c\}$ &$l/q$ \\
    \hline
    $\ydiagram{1,1,1}$ &6& 3 & 3 &2\\
    \hline
  \end{tabular}
  \caption{Value of each indicators in $SU(6)$  $\ydiagram{1,1,1}$\label{tab:SU6 の値}}
\end{table}
\subsubsection{Chiral symmetry }
This theory has classical global $\mathrm{U}(1)$ symmetry;
\begin{equation}
    \psi \mapsto e^{i\alpha} \psi, \quad \bar{\psi} \mapsto e^{-i\alpha} \bar{\psi}, \qquad (\alpha:\text{constant}).
\end{equation}
In quantam theory, however, this symmetry is broken by ABJ anomaly, which is given by the transformation of path integral measure such as 
\begin{equation}
    \label{change of integral mesure}
  \int \mathcal{D}\psi \mathcal{D}\bar{\psi} \mapsto \int \mathcal{D}\psi \mathcal{D}\bar{\psi} e^{i\alpha l \nu},\quad \nu := \frac{1}{8\pi ^2} \int \tr (F \wedge F),
\end{equation}
where $\nu \in \mathbb Z$ is the instanton number. Hence, eq.(\ref{change of integral mesure}) is even invariant when $\alpha = 2\pi n/l, \: n \in \mathbb Z $. It corresponds to the $\z l ^\chi \subset U(1)$;
\begin{equation}
    \psi \mapsto e^{2\pi i n /l } \psi, \quad \bar{\psi} \mapsto e^{-2\pi i n / l} \bar{\psi}, \qquad (n \in \mathbb{Z}),
\end{equation}
which let us call discrete chiral symmetry.

\subsubsection{Center symmetry}
In general,  the center symmetry $\z N$ is partially or completely broken due to the existence of matter fields as
\begin{equation}
    \z N \mapsto \z {q = \gcd (c,N) } \in \z N,
\end{equation}
except the case that fermions is in adjoint representation \cite{Tanizaki:2018wtg}.
We denote the N-ality (the number of boxes of Young tableau) c, 
then the unbroken center symmetry is given by $\z q \in \z N$, where $q = \gcd (c,N)$. Therefore we can rewrite N and c with q as $N=N_0 q, \: c = c_0q$.

In the case of $N=6$, N-ality is $c=3$, so that  $q =\gcd (c,N) = 3$.

%%%%%%%%%%%%%%%%%%%%%%%%%%%%%%%%%%%%%%%%%%%%%%%%%%%%
%%%%%%%%%%%%%%%%%%%%%%%%%%%%%%%%%%%%%%%%%%%%%%%%%%%%
%%%%%%%%%%%%%%%%%%%%%%%%%%%%%%%%%%%%%%%%%%%%%%%%%%%%
\begin{comment}
    \subsubsection{\texorpdfstring{$\z 2$}{} redundancy}
\label{z2 redundancy }
Let us denote the component $\omega = e ^{i \frac{2 \pi}{6}n} \in \z 6$, where $n \in \mathbb Z$.

As discussed above, this system has $SU(6)/\z {q=3}^c \times \z 6^\chi$ symmetry.
Then we can also consider the transformations operating Weyl fermions as follows.
Weyl fermions in the system are transformed under 0-form $\z {N=6} ^{c (0)} \subset SU(6)$ as $\psi \mapsto \omega ^3 \psi $, since $\psi$ is in third-order totally antisymmetric representation; it's N-ality is 3. This transformation is equivalent to $\z 2 ^{c (0)} \subset SU(6)/ \z {q=3}$  %\subset \z {N=6} ^{c (0)} $.

It is obvious that this $\z 2 ^{c (0)} = SU(6)/ \z {q=3}$ transformation is overlapping with $\z 2 ^\chi \subset \z 6^\chi$. Therefore the total symmetry is given by eq.(\ref{gauge group}).

\end{comment}

%%%%%%%%%%%%%%%%%%%%%%%%%%%%%%%%%%%%%%%%%%%%%%%%%%%%
%%%%%%%%%%%%%%%%%%%%%%%%%%%%%%%%%%%%%%%%%%%%%%%%%%%%
\subsection{Chiral symmetry breaking}
\label{Chiral symmetry breaking}
As mentioned above, this system has 1-form center and 0-form chiral mixed anomaly, $ \z {l} ^{\chi(0)}-[\z q^{c{(1)}}]^2$, which leads to the spontaneous discrete chiral symmetry breaking,
\begin{equation}
    \z l ^\chi \mapsto \z {l/q =2} ^\chi.
\end{equation}
Therefore there are three distinct vacua related by the broken elements $\z l / \z q \simeq \z {l/q =2}$.
Note that chiral symmetry breaking in this model is guaranteed mathematically by using the fact derived in \cite{Cordova:2019bsd}\cite{Cordova:2019jqi} assuming the theory is gaped.

It is also remarkable that fermion bilinear condensate expected as the order parameter of the symmetry breaking is always identically zero i.e.,

\begin{equation}
 \langle \psi\psi \rangle = 0,
\end{equation}
%\begin{equation}
 %< \psi\psi > := \epsilon^{\alpha\beta} \psi^I_\alpha \psi ^J_\beta B_{IJ} = 0,
%\end{equation}
due to the Fermi-Dirac statics.
However, there has been some discussion about this, which implies the possibilities of the existence of non-vanishing fermion bilinear (See \cite{Bolognesi:2019fej} for details.).

%%%%%%%%%%%%%%%%%%%%%%%%%%%%%%%%%%%%%%%%%%%%%%%%%%%%
%%%%%%%%%%%%%%%%%%%%%%%%%%%%%%%%%%%%%%%%%%%%%%%%%%%%
\section{Gauging the total symmetry}
\label{Gauging the chiral symmetry}
%\subsection{'t Hooft anomaly from Anomaly polynomial }
The mixed anomaly for U$(1)$ and $\z q$ described above was obtained by gauging part of the gauge symmetry as mentioned in Sec.\ref{the symmetries}.
In this section, our aim is to gauge all gauge symmetries in eq.(\ref{gauge group}) to extract all of the 't Hooft anomaly information available at UV theory.
Fortunately, in our system, it can be achieved by the Stora-Zumino procedure \cite{Manes:1985df}. This property is unique to this model.
So this is not always true in any model.

\subsection{'t Hooft anomalies via Stora-Zumino procedure  }
\label{'t Hooft anomaly by Stora-Zumino procedure}
Here, we compute the 't Hooft anomaly via The Stora-Zumino procedure\cite{Zumino:1983ew}.
Ref.\cite{Bolognesi:2019fej} also evaluated it but took only the linear term of \zn 6 gauge field, $A_\chi^{(1)}$.

The 6-dimensional anomaly polynomial is given by
%%%%%%%%%%%%%%%%%%%%%%%%%%%%%%%%%%%%%%%%%%%%%%%%%%%%%
\footnote{
We should consider the anomalies associated with gravitation, too. However, we figure out that this model has no such anomalies by the analysis in \cite{Hsieh:2018ifc}.
Therefore we ignore the terms associated with gravitation in the discussion.}
%%%%%%%%%%%%%%%%%%%%%%%%%%%%%%%%%%%%%%%%%%%%%%%%%%%%%
\begin{equation}
    \mscr{A}_6 =  \int \frac{2\pi}{3! (2\pi)^3} \Tr _c \[ \mcl {R}(\Tilde{F}-B_q^{(2)}) + dA_\chi ^{(1)}  \]^3  %+ \Tr _c \( \mathrm{d} A_\chi^{(1)}  \) 
    %\dim R \; \mathrm{d} A_\chi^{(1)} \wedge \frac{1}{24}\frac{1}{8 \pi ^2} \tr \[ R \wedge R \]
    %\frac{2\pi}{3! (2\pi)^3} \int  \Tr _c \[ dA_\chi ^{(1)} \wedge \frac{1}{8} \tr \( R \wedge R \)  \] ,
\end{equation}
which leads to the 5-dimensional SPT action;
\begin{equation}
    {S}_{\tx{SPT}} = \int  A_\chi^{(1)} \wedge \[  \frac{2\pi}{3! (2\pi)^3} \( 3l \; \tr (\Tilde{F}-B_q^{(2)})^2 + \dim R \; (dA_\chi ^{(1)})^2 \) + \frac{\dim R}{24} p_1(M_4)   \].
\end{equation}

This is obviously invariant under the 1-form center transformation, $\z {q=3}^{c(1)}$.
Under the chiral transformation, $\z {l=6}^{\chi (0)}$, the SPT action gives the 't Hooft anomalies;
\begin{align}
\label{delta SPT}
    \delta {S}_{\tx{SPT}} &= \frac{2\pi}{6}  \int \[ \frac{2\pi}{3! (2\pi)^3} \( 3l \; \tr (\Tilde{F}-B_q^{(2)})^2 + \dim R  (dA_\chi ^{(1)})^2\)  \]\notag\\
    &= \frac{2\pi}{6}  \int \[ \frac{2\pi}{3! (2\pi)^3} \( 3l \; \tr (\Tilde{F})^2 -3lN  (B_q^{(2)})^2 + \dim R \; (dA_\chi ^{(1)})^2 \)  \]
    %&= \frac{2\pi}{3! (2\pi)^3} \int \frac{2\pi}{6}  \[ 3l \tr (\Tilde{F})^2 -3lN  (B_q^{(2)})^2 + \dim R \( (dA_\chi ^{(1)})^2 - p_1(M_5)  \)  \] ,
\end{align}
where we use the constraint, $\tr \(  \Tilde{F}-B_q^{(2)}\) = 0$ in the second line.
The first term in eq.(\ref{delta SPT}) is trivial under the $\frac{SU(6)}{\z 3^c} \times \z 6^\chi$ bundle.
The second term leads to the chiral and center mixed anomaly;
\begin{equation}
\label{anomaly Zl-ZcZc}
     \mcl {A}_{[\z 6 ^{\chi (0)}]-  [\z 3 ^{c (1)}]^2 } \equiv -\frac{2\pi}{3! (2\pi)^3} \int \frac{2\pi}{6} 3lN \; (B_q^{(2)})^2 \in -\frac{2}{3} \cdot 2\pi\mathbb Z,
\end{equation}
which is consistent with \cite{Yamaguchi:2018xse,Bolognesi:2019fej}.
The third term corresponds to the pure discrete chiral anomaly;
\begin{equation}
 \label{anomaly ZlZlZl}
    \mcl {A}_{[\z 6 ^{\chi (0)}]^3 }\equiv \frac{2\pi}{3! (2\pi)^3} \int \frac{2\pi}{6} \dim R \; (dA_\chi ^{(1)})^2 \in \frac{1}{9} \cdot2\pi\mathbb Z,
\end{equation}
which coincides  with the result in the computation  in\cite{Hsieh:2018ifc,Garcia-Etxebarria:2018ajm} as  we will see in Sec.\ref{Pure discrete chiral anomaly}.
%And the fourth term is the gravitational mixed anomaly,
%\begin{equation}
%\label{anomaly zl gravity}
%    \mcl {A}_{[\z 6 ^{\chi (0)}]-[grav.]^2 } \equiv  \frac{2\pi}{3! (2\pi)^3} \int \frac{2\pi}{6} \dim R \; p_1(M_4) \in  \frac{20}{3} \cdot 2\pi\mathbb Z.
%\end{equation}

Let us remind that non-perturbative anomalies such as $[\z 6 ^{\chi (0)}]^3,  $ have sometimes nontrivial values which cannot be captured by the anomaly polynomial.
which implies that eqs.\eqref{anomaly ZlZlZl}  might be incorrect.
Therefore, in the next section, we see whether our results is correct using the result of rigorous computation of $\eta$ invariant.
Then we realize that the non-perturbative anomalies in our system is precisely evaluated via the Stora-Zumino procedure.

%%%%%%%%%%%%%%%%%%%%%%%%%%%%%%%%%%%%%%%%%%%%%%%%%%%%
%%%%%%%%%%%%%%%%%%%%%%%%%%%%%%%%%%%%%%%%%%%%%%%%%%%%
\subsection{Non-perturbative chiral anomalies}
\label{Pure discrete chiral anomaly}
%\paragraph{Pure discrete chiral anomaly}
Analysis of the $\eta$ invariant provides the result of 'pure discrete $\z n ^{(0)} $ gauge anomaly' under the symmetry transformation of Spin$(4) \times \z N$, as follows \cite{Hsieh:2018ifc,Garcia-Etxebarria:2018ajm}
\begin{equation}
  \label{discrete anomaly free condition}
  \mcl {A}_{[\z n ^{(0)}]^3 }\equiv (N^2 + 3 N + 2)  \( \sum_L s_L^3 -  \sum_R s_R^3 \)  \mod 6n,
\end{equation}
where $ s_L, s_R $ are the $\z n$ charges of fermions. 
We apply the eq.(\ref{discrete anomaly free condition}) to our system, then we find that the pure discrete chiral anomaly $[\z 6 ^{\chi (0)}]^3 $ is valued in $1 \mod 9$.
Surprisingly, the result is agree with $\mcl {A}_{[\z 6 ^{\chi (0)}]^3 }$, eq.(\ref{anomaly ZlZlZl}).

%%%%%%%%%%%%%%%%%%%%%%%%%%%%%%%%%%%%%%%%%%%%%%%%%%%%
%%%%%%%%%%%%%%%%%%%%%%%%%%%%%%%%%%%%%%%%%%%%%%%%%%%%

\section{IR effective theory}
In this section, We derive the one of the possible effective IR actions with the idea of Wess-Zumino-Witten term\cite{Witten:1982fp}\cite{Zumino:1983ew}, which imposing to recapitulate all anomalies in high-energy region
\footnote{
Our procedure is very similar to \cite{Anber:2020xfk,Anber:2021iip} which match the mixed anomalies arising UV region.

}.

In order to achieve our goal, here, let us assume following;
 \begin{itemize}
     \item The system is in a confinement  phase in IR scale.
     \item The system is not constructed by CFT.
 \end{itemize}
Under these assumptions, our system has to break the chiral symmetry (Sec.\ref{Chiral symmetry breaking}).

\subsection{Effective action for mixed anomaly}
\label{Effective action}
The WZW term states that the contribution of NG fields to the anomaly is expressed as the difference between the shifted CS term of bare gauge fields $(A_h,A)$ and that of $((A^{U^{-1}})_h,A^{U^{-1}})$.
In other wards, we can choose the NG fields, so that the shifted CS term of dressed gauge fields $A^{U^{-1}} $ is gauge invariant.

Of course we know that the no NG boson arises since broken chiral symmetry is discrete. However, we can expect that some composite scalar fields such as $\psi\psi$ or $\psi\psi\psi\psi$ which can be interpreted as the order parameter of the chiral symmetry breaking exist in low energy region reproducing the UV anomalies.

From this view point, let us assume that there exists the composite scalar field $\phi$ with charge $Q$ under the $\z 6 ^{\chi(0)}$ transformation, which satisfies 
\eqref{z6 trsf. of phi}.
Then, we can construct the chiral invariant action, which corresponds to the shifted CS term $\tilde \omega_5^{(0)} ((A^{U^{-1}})_h,A^{U^{-1}})$, with $\Phi$ in 5-dimensional manifold N
\begin{align}
\label{Omega5}
    \Omega_5 = &  \int_N (d \Phi - A_\chi^{(1)}) \wedge \[  \frac{2\pi}{3! (2\pi)^3} \( -3lN(B_q^{(2)})^2 + \dim R \; (dA_\chi ^{(1)})^2 \)  \] \notag \\
    & +{\frac{q}{2\pi } \int _N d\phi\wedge db^{(3)}}
\end{align}
where $\Phi \equiv \frac{\phi}{Q}$ and 
\begin{align}
%\label{definition of phi}
\label{z6 trsf. of phi}
     \quad \z 6^{\chi(0)}: \Phi \mapsto \Phi + \frac{2\pi}{6}, \quad \phi \mapsto \phi + Q \frac{2\pi}{l}.
\end{align}
The last term in eq.(\ref{Omega5}) is the Lagrange multiplier and 
$\oint _M  db^{(3)}  \in 2 \pi \mathbb Z $ .
Therefore the effective IR action is 
\begin{equation}
  \label{IR action}
      S_{IR}= \int_N  \Phi \wedge \[  \frac{2\pi}{3! (2\pi)^3} \( -3lN(B_q^{(2)})^2 + \dim R \; (dA_\chi ^{(1)})^2 \)   \]
      +  {\frac{q}{2\pi } \int _M \phi \wedge db^{(3)} }.
  \end{equation}
  Note that we can construct the effective action by just one scalar field which reproduce the mixed 't Hooft  anomaly (\ref{anomaly Zl-ZcZc}) in the UV energy scale.

%%%%%%%%%%%%%%%%%%%%%%%%%%%%%%%%%%%%%%%%
%%%%%%%%%%%%%%%%%%%%%%%%%%%%%%%%%%%%%%%%
\subsection{The nontrivial topological term for chiral pure anomaly}
On the other hand, in the case of the self-anomaly,
it is very non-trivial which degrees of freedom reproduce the self anomaly.
 This is my ongoing work.
 First, the topological term given by Stora-Zumino procedure is actually, ill-defined mathematically.
 It can bee seen by deforming to the cochain form.
 And the well-defined topological term in cochain form is proposed by Wan and Wang as follows \cite{Wan:2018bns},
\begin{align}
  \eta_{\text{chiral} } = \beta_9 \(\beta_3 A_3 \cup \beta_3 A_3 \),
\end{align}
where $A_3$ is given by docomposing $A_\chi$ as $A_\chi = A_2 + A_3,\, A_2 \in \z 2, A_3 \in \z 3$, and we negrect $A_2$ because $\z 2$
symmetery has no anomaly.
$\beta_3, \beta_9$ are Bockstein homomorphism satisfying
\begin{align}
  \beta_9 &: H^n(-, \mathbb{Z}_3) \to H^{n+1}(-, \mathbb{Z}_9)\\
  \beta_3 &: H^n(-, \mathbb{Z}_3) \to H^{n+1}(-, \mathbb{Z}_3).
\end{align}

Now it turns out what we have to do is to find the 4-dimensioinal term which compensate with this topological term.
This is  just problem we are working on now.

In the very naive discussion, this topological term may be similar to the CS term.
Indeed, 
under the assumption that the topological term forms like CS term, \cite{Delmastro:2022pfo,Kaidi:2023maf} succeeded in predicting even more.

If this is true, the self-anomaly might also be matched by the scalar field $\phi \in Z_3$. Then the vacuum structure is very simple.
Another possibility is that some degrees of freedom on the domain wall, which is inserted between degenerate vacua, match the anomaly.

\section{Summary}
This study identifies two anomalies in the model under consideration.
 The \textit{mixed anomaly} can be matched by the presence of $\phi \in \mathbb{Z}_3$. 
 On the other hand, the nature of the \textit{discrete chiral anomaly} remains partially unresolved. 
 Although a corresponding topological term is known, the associated four-dimensional degrees of freedom 
 remain unclear. Based on insights from previous studies, 
 it is anticipated that this topological term can be reformulated in a form similar to the Chern-Simons term. 
 If this is achieved, the discrete chiral anomaly can also be matched by $\phi \in \mathbb{Z}_3$. 
 This finding implies that the vacuum structure of the model may be fully constructed using $\phi \in \mathbb{Z}_3$.
The above discussions provide a crucial foundation for future efforts to realize this model on the lattice.

The work of H.W. was supported in part by JSPS KAKENHI Grant-in-Aid for JSPS fellows Grant Number 24KJ1603. 
The work of T.O. was supported in part by JSPS KAKENHI Grant Number 23K03387.
The work of T.Y. was supported in part by JST SPRING, Grant Number JPMJSP2138.
\appendix

\section{Cancellation of \texorpdfstring{$\z 2$}{} symmetry}
\label{Cancellation of z 2 symmetry}
It is important to identify the relevant total symmetry of the system to achieve our goal.
First, let's consider the total group, eq.(\ref{gauge group}), again.
Intuitively, it would seem that $\z 2$ have also to be further gauged, but actually we can find that the relevant gauge group is the the third term in eq.(\ref{gauge group again});
\begin{equation}
\label{gauge group again}
    G =  \frac{SU(6)\times \z 6^\chi}{\z 6}\cong \frac{SU(6)/\z 3^c \times \z 6^\chi}{\z 2} \sim \frac{SU(6)}{\z 3^c} \times \z 6^\chi.
\end{equation}
The third equal "$\sim$" is justified because the 2-form background gauge field of $\z 2$ can be vanished by the 1-form gauge transformation.
Let us derive this fact.
First, we consider the $\z 6 ^\chi / \z 2  $ bundle. 
Its cocycle condition is twisted;
\begin{equation}
    \omega ^{n_{ij} + n_{jk} + n_{ki}} = \omega ^{3 b_{ijk}} = \exp{\(\frac{2\pi i}{2}b_{ijk}\)},
\end{equation}
where $\omega ^{n_{ij}} = \exp{\(\frac{2\pi i}{6} n_{ij} \)} \in  \z 6 ^\chi$ is the transition function  on $U_{ij} = U_i \cup U_j $ and $b_{ijk} \in \z 2$ corresponds to the 2-form background gauge field of $\z 2$ .
This twisted cocycle condition is compensated with that of $\frac{SU(6)/\z 3}{\z 2}$ bundle.
The transition functions  $\omega ^{n_{ij}}$ is changed under $\z 2$ gauge transformation as 
\begin{equation}
    \omega ^{n_{ij}} \mapsto \omega ^{n'_{ij}} = (-1)^{m_{ij}} \omega ^{n_{ij}} = \omega ^{n_{ij} + 3 m_{ij}}
\end{equation}
It leads 
\begin{equation}
     \omega ^{n'_{ij} + n'_{jk} + n'_{ki}} =  \omega ^{n_{ij} + n_{jk} + n_{ki} + 3 {m_{ij} + m_{jk} + m_{ki}}} = \omega ^{3 \(b_{ijk} + m_{ij} + m_{jk} + m_{ki} \)} \equiv \omega ^{3 b'_{ijk}},
\end{equation}
where $ b'_{ijk} = b_{ijk} + m_{ij} + m_{jk} + m_{ki}$.
Then, if we chose $m_{ij} = n_{ij}$, 
\begin{equation}
    \omega^{n'_{ij} + n'_{jk} + n'_{ki}} = \omega ^{4 \cdot 3b_{ijk}} = 1.
\end{equation}
Therefore, we don't have to gauge 1-form $\z 2$ symmetry, so that the gauge symmetry group we want is the last term in eq.(\ref{gauge group again}).

This fact is mathematically rigorous.
In other words, the cohomology of $\z 2$ is trivial.
This shows that no anomaly associated with symmetry $\z 2$ arises in the anomaly analysis.
Thus, the only 't Hooft anomaly that arises from this total symmetry is the one concerning $\z q ^{c(1)}$ and $\z l ^{\chi(0)}$.

If we introduce the $n_f $flavor symmetry, above discussion gets quite complicated.
The total symmetry is given by 
\begin{equation}
    \frac{ \frac{\mathrm{SU}(6) }{\z 3^c} \times \frac{\mathrm{SU}(n_f) \times \z {n_f l} }{\z {n_f}}  }{\z 2}.
\end{equation}
To consider whether the redundancy $\z 2$ can be cancelled, in short, what we have to see is that whether the following equation is  satisfied;
\begin{equation}
    \frac{\mathrm{SU}(n_f) \times \z {n_f l} }{\z {n_f}} \cong \frac{ \frac{\mathrm{SU}(n_f) \times \z {n_f l} }{\z {n_f}}  }{\z 2} \times \z 2.
\end{equation}
In terms of this equation, we can realize that the case of $n_f =1$ is satisfied as discussed.

%%%%%%%%%%%%%%%%%%%%%%%%%%%%%%%%%%%%%%%%
%%%%%%%%%%%%%%%%%%%%%%%%%%%%%%%%%%%%%%%%
\bibliographystyle{utphys}
\bibliography{refer}

\end{document}